**Nuclear Power and the World's Energy Requirements.**

V. Castellano, R. F. Evans and J. Dunning-Davies,
Department of Physics,
University of Hull,
Hull HU6 7RX,
England.

J.Dunning-Davies@hull.ac.uk

**Abstract.**

The global requirements for energy are increasing rapidly as the global population increases and the under-developed nations become more advanced. The traditional fuels used in their traditional ways will become increasingly unable to meet the demand. The need for a review of the energy sources available is paramount, although the subsequent need to develop a realistic strategy to deal with all local and global energy requirements is almost as important. Here attention will be restricted to examining some of the claims and problems of using nuclear power to attempt to solve this major question.

## Introduction.

One of the more important factors involved when determining the overall development of a country is its energy consumption. It is undoubtedly the case that this factor provides a major difference between the so-called developed and under (or less) developed countries of the world. During the post-war period, the rapid development of the economies of the Western World was linked closely to oil, and possibly still is. Oil was used for a wide variety of purposes, for electricity production, for transport, as well as in the growth of the entire petrochemical industry. However, the oil crises of 1973 and 1979 produced a change in attitude and the main change was in the effort employed to make the West less vulnerable to the power of the major oil providers. This change did not affect the developed world too drastically, but the under-developed countries fared less well and many plunged even further into debt. With the population of the under-developed world being larger than, and increasing faster than, the population of the West, it seems the situation can only deteriorate.

In 1999, the United Nations announced that the world's population had reached six billion, a mere twelve years after reaching the five billion mark. It is predicted that the figure of seven billion will be achieved between 2011 and 2015. The actual outcome depends crucially on the situations in China and India, the two most highly populated countries of the world, which between them are home to some 38% of the world's population. At present, China has 1,281 million inhabitants, India 1,050 million; but by 2050 it is predicted that India's total population will have overtaken China's total, having approximately 1,600 million inhabitants compared with China's predicted figure of 1,400 million. The reason these two countries are so important in any consideration of energy needs is because both are counted in the group of under-developed countries and, while at present the less developed countries account for approximately 80% of the world's population, by 2050 it is predicted that that figure will have risen to about 86% [1]. These figures are of vital importance when considering global energy requirements since, at present, the under-developed part of the world uses far less energy per head of population than does the developed part. It is estimated, as shown below, that twelve times as much energy per person is used in the developed countries as compared with the under-developed ones. However, that situation is changing rapidly as the under-developed countries desperately strive to catch up with the rest. A further problem, which could increase in the future, is that much of this energy is provided by the combustion of fossil fuels [2], resulting in the production of large quantities of $CO_2$, $SO_x$ and $NO_x$, with the attendant problems of increased global warming and acidification of rain [3].

It has been estimated [2] that the present energy consumption of the world is in the region of $2 \times 10^{20}$ Joules per year, which equates to a rate of working of something of the order of $0.63 \times 10^{13}$ Watts. With the world population being approximately six and a quarter billion, it follows that each person accounts for about 1 kW. However, this figure totally ignores the fact that the approximately 20% of the world's population, or about one and a quarter billion people, inhabiting the developed world consume roughly 75% of the energy produced, or about 3.78 kW per head of population. As a complete contrast, the five billion people of the under-developed world are able to call on the remaining 25% of the energy produced in the world and this corresponds to a mere 0.315 kW per head of population.

For good and understandable reasons, the underdeveloped sections of the world are attempting to catch up with the more developed areas. Achieving this, would improve so many aspects of life for so many people; in particular, health should improve and life expectancy increase. However, any such successful modernisation would necessarily include an enormous increase in energy consumption. At the moment, if all people in the world consumed the same amount of energy as those in the developed countries, the total amount of energy required would be $\approx 2.36 \times 10^{13}$ Watts. This nearly quadruples the present level of energy consumption, leading to a projected annual energy requirement of $\approx 7.5 \times 10^{20}$ Joules. Such a requirement, when it arises, may be met only by increased availability of energy. This extremely rough estimate ignores the fact that the world's population is increasing quite rapidly and assumes that any technological advances will not cause increased drains on energy resources. The problem of satisfying the world's energy needs is a major one and needs to be addressed urgently since the solution cannot be simply increasing the present methods of energy provision. This is so because the energy sources relied on now represent a finite energy reservoir and also some of the thermodynamic implications of present practices need examining if a clean environment is to be produced for future generations.

**Traditional sources of energy.**

The reserves of fossil fuels are known to be finite and, even at the current level of usage, their life-times are fairly small. In fact, it might be noted that already in 1999 and the first quarter of 2002, the total world demand for oil exceeded the total world supply [4]. These two cases may be merely blips in the statistics but, nevertheless, sound a warning as far as dependence on oil is concerned. Coal, on the other hand, presents different problems. The stocks are diminishing rapidly, the cost of extraction in some cases is increasing and, like oil, it contributes considerably to the planet's environmental problems when used as a fuel. Another major source in the West is provided by natural gas which has the advantage of not producing high quantities of $CO_2$ when burnt, but its stocks are strictly limited. Furthermore, when the above population figures and the relative sizes of the developed and under-developed sections of the world are noted, it is seen that the energy requirements of the world are certain to rise drastically in the near future. This means that, even allowing for the possible discovery of new resources, fossil fuels will be unable to provide the world with sufficient energy for any significant length of time. It might be noted also that fossil fuels are used extensively in both the pharmaceutical and petrochemical industries, where substitutes prove expensive alternatives.

The unfortunately named 'renewable' energy sources, although quite numerous and varied, are unlikely to be able to contribute significantly more than about 20% of future total energy requirements [5]. These sources include geothermal energy, solar energy, wind power and wave power. Numerous though these may seem, it remains extremely unlikely that, taken together, these could combine to satisfy the world's future energy needs, especially if increased demand is accompanied by a decrease in the availability of fossil fuels as seems likely. All these sources of energy must surely have an important rôle to play, but it should always be remembered that while these sources are termed 'renewable', and although they truly seem non-decreasing, they too represent finite sources ultimately; - the second law of thermodynamics would allow nothing else!

It is well-known that, in the regions of the earth not too far from the surface, there is a temperature gradient of roughly 30K/km. In some places, the higher temperatures below the suface lead to geysers and other phenomena. However, the heat distribution is not uniform, with the temperature gradient being much greater in some places than others. A geyser is formed if water accumulates deep down where it is turned into steam which builds up in pressure before breaking through the earth's surface. Some of these naturally occurring phenomena have been harnessed to provide superheated steam which, in turn, may be used to provide power. Such plants may well make a useful contribution to energy needs but they are unlikely to prove significant globally and so geothermal energy sources are not likely to make any worthwhile impact as far as global energy needs are concerned.

Wind and solar power, the two major regenerative sources, face the major problem of requiring a substantial portion of the earth's surface to provide the required energy. It has been speculated [2] that, at some time in the future, if the reliance on these two sources was increased, that portion could be 10% or more. What is more, such land surface would have to be in carefully chosen, appropriate places; possibly in the tropics for solar power or in known windy regions for wind power. There would also be associated transmission problems but, possibly more importantly, although wind and solar power sound attractive to many people initially, as soon as the amount of land to be committed to such schemes became known, it is likely that social objections would be raised quite forcibly. Further, both sources would be unable to guarantee actual production at any particular time and so substantial high power storage facilities would be needed and, as yet, no such facility exists. It has been estimated [2] that these two sources could not provide more than about 20% of Britain's energy requirements and possibly less for some other northern countries. These two sources must be remembered, however, as long term possibilities for at least helping provide for the world's energy needs.

The harnessing of wave power presents its own set of seemingly enormous engineering problems and, so far, it seems there has been little progress in solving the practical problems of energy conversion associated with this form of power. However, looming over everything is the shear power of the sea. It will be a truly tremendous feat of engineering to produce a device which is able to harness the power of the sea for our energy needs; a device that is robust enough to withstand major storm conditions and yet delicate enough to operate efficiently in conditions of relative calm. Any deployment of collectors for such a system would inevitably affect shipping and it is doubtful that any system would satisfy the worlds' total energy needs, at least not in the near future. However, this is certainly another potential source not to be forgotten.

Another source of energy, particularly popular in some parts of the world, is biomass. However, this source presents a big danger because its abuse could accelerate the world deforestation process. This source is another which should not be termed 'renewable' since, at present rates, for every ten hectares cut down, only one is being replanted. Another disadvantage with this fuel is that it provides another source of contamination of the atmosphere.

Other potential sources, such as ocean thermal power and the hydraulic resource, as well as further details of the above-mentioned sources have been discussed elsewhere [2]. It seems, unfortunately, that wave power, biomass, geothermal energy and tidal sources have all been found lacking when it comes to providing for the worlds' future probable energy needs; they provide insufficient power for present, leave alone future, purposes. This leaves the fossil fuels, which are slowly but surely disappearing, and nuclear power.

**Nuclear power.**

At present, nuclear fission reactors provide a significant proportion of the world's energy, with approximately four hundred nuclear plants being in operation and producing of the order of 17% of the world's electricity. High concentrations of these plants are to be found in the U.S.A., Japan and Europe. However, once again there is reliance on a finite source of fuel, uranium; although, in terms of power production potential, resources are much greater than is the case for fossil fuels. In many ways, as far as the projected time for which mankind might survive is concerned, one major sustainable method of energy production is provided by fast breeder reactors. In these reactors, under appropriate conditions, the neutrons given off by fission reactions can 'breed' more fuel from otherwise non-fissionable isotopes. The most commonly used reaction for this purpose is by obtaining plutonium $^{239}$Pu from non-fissionable uranium $^{238}$U. The term 'fast breeder' refers to the situations where more fissionable material is produced by the reactor itself. This latter situation is possible because uranium $^{238}$U is many times more abundant than fissionable uranium $^{235}$U and may be converted into plutonium $^{239}$Pu, which may used as fuel, by the neutrons from a fission chain reaction. Attractive though such reactors may appear at first, they prove to be extremely expensive, largely due to important safety concerns surrounding the use of molten metals to remove the huge quantities of heat produced and to the fact that the fuel is highly radioactive plutonium. However, nuclear power always raises great worries with many people on at least two counts: firstly, there is always worry over a possible accident occurring, and secondly there is worry over the disposal of any radioactive waste. Countries such as the UK and Japan reprocess a proportion of the waste for use in weapons and medical facilities. However, this is expensive and time consuming and should be viewed as a form of recycling, rather than waste 'disposal'. In countries such as France and the USA, the majority of the waste is stored in water tanks on the actual sites of the nuclear fission reactors. This has led to a huge build-up, over the past fifty years, of a substantial stockpile of highly radioactive waste. This has prompted the need to find essentially permanent storage facilities for the material and, for example, the American government is presently having such a storage facility constructed at Yucca mountain in Nevada. This proposed facility is proving an enormously expensive exercise as reported in the National Geographic [6].

The big growth in the use of nuclear power came approximately thirty years ago and was probably due to the oil crises of the seventies. As soon as the price of oil returned to normality, however, nuclear energy ceased being competitive, mainly because of the high costs associated with basic nuclear technology. These costs are recoverable in the long term and proof of that claim is provided by realising that in 2002, the cost in cents per kWh of electric generation was 1.76 for nuclear power, 1.79 for coal, 5.28 for oil and 5.69 for gas; where these costs cover fuel, operation and maintenance, but not capital costs [7]. Hence, nuclear power was able to undercut other forms of energy generation and so should, in the longer term, be capable of recovering the initial capital outlay without losing the lowest position on the cost scale. It is always worth remembering also that, while there are drawbacks associated with the use of nuclear power (drawbacks which are outlined above), its use does not produce the dangerous gases which are polluting the atmosphere and causing acid rain. These may seem small points but everything needs to be taken into account when attempting to assess the provision of the worlds' future energy requirements.

**Conventional methods for the disposal of radioactive waste.**

Radioactive material that cannot be utilised directly in other processes is designated nuclear waste and most nuclear processes produce amounts of such waste. Long term solutions for its safe disposal have been sought for many years but, even today, few suggested solutions have been implemented. There are, in fact, several categories of waste but here attention will be restricted to a consideration of methods for disposing of so-called 'high level' waste.

Modern conventional nuclear reactors (advanced gas reactors and pressurised water reactors) use enriched uranium fuel as a heat source. This is made from natural uranium ore which typically contains about 0.7% uranium $^{235}U$, enriched to between two and three per cent, depending on the requirements of the particular reactor. This leaves a large amount of uranium $^{238}U$ with a reduced concentration of uranium $^{235}U$; this is classed as 'medium level' radioactive waste. The enriched fuel is then compacted into fuel rods as $UO_2$, ready for use in a reactor core. When exposed to 'thermalised' neutrons, the uranium $^{235}U$ undergoes stimulated fission, leading to the production of a great variety of radioactive by-products, stored in the fuel rods. Once the concentration of uranium $^{235}U$ drops below about 0.9%, the fuel rod is classed as 'spent' and a new rod replaces it. The 'used' fuel rods produce a considerable amount of heat due to their high level of radioactivity - approximately $3 \times 10^8$ times that of a new fuel rod - and are stored typically in ten metre deep water pools on site for at least twelve months. This storage is to allow them to cool and for their radioactivity to decrease to a safer level. These 'cool' rods are then felt safer to transport and may be sent either to a reprocessing plant where useful products such as plutonium and the remaining uranium may be extracted or, more usually, may be moved to a large, longer-term storage facility.

The reprocessing of the fuel rods is achieved by cutting them up and dissolving them in nitric acid. This releases most of the gaseous fission products into solution; the exception being the noble gases. Most of the radioactivity in the spent fuel rods ($\approx 76\%$) originates from the fission products, except plutonium. Since the plutonium and remaining uranium are of use, they are removed from the solution chemically, leaving the highly radioactive waste in solution. This solution is then stored for a number of years before being evaporated and vitrified into glass blocks for long-term storage. This process, although seemingly efficient, in that the final waste material contains about 97% of the waste fission products, produces a large amount of low and intermediate waste which must be disposed of also. However, once waste is in the form of vitrified waste or cool fuel rods, it may be 'disposed' of either by being placed sufficiently out of harms way so that it requires no more monitoring or alternatively by being 'neutralised' by conversion to a harmless substance.

At present, the most popular method is to store the waste deep underground in very stable geological sites so that, by the time the waste leaks out, it is of no danger to life on earth. Such sites are required to be such that the waste may be safely stored for of the order of 400,000 years. One major problem with this, however, is that there is little evidence to support the supposition that the containers designed for the task would themselves survive for such a long time. There is also a great deal of controversy over the levels of seepage of radioactive elements from the stored waste, since predictions over such a long period of time are fraught with inherent uncertainties.

It is interesting, and possibly instructive, to consider data from what amounts to a natural uranium reactor, which provides a precedent for radio-isotope distribution over a very long time scale. A recently discovered site in West Africa had an unusually low concentration of uranium $^{235}U$ within the uranium ore. The only way it is felt this can be explained is if a significant proportion of the original uranium $^{235}U$ underwent fission. The area of land concerned is saturated with water which would provide a moderator capable of thermalising the neutrons. If the concentrations of uranium $^{235}U$ were sufficiently high, it is perfectly possible for a natural fission reactor to operate. Indeed, the concentrations of radioactive products indicate that this natural reactor operated approximately 1.8 billion years ago. When measurements were taken to see how far the metallic radioactive products had travelled in that time, it was found to be less than a metre from the original reactor site. Although the data is specific to the site in question, it does suggest that the level of transport of waste may be insignificant as far as the human race is concerned.

Another method of dealing with radioactive waste, which is under consideration at present, is the conversion of the waste into less dangerous materials, usually through high intensity neutron bombardment. The idea is currently still at the development stage but its main disadvantage is the low volume of waste that can be practically converted in this way.

**An alternative method for disposal of high-level radioactive waste.**

An alternative method for disposing of high-level radioactive waste has been proposed recently by Santilli [8]. It is a form of neutralisation but does not use the conventional methods currently being researched. Indeed, classical formulations of quantum chemistry and nuclear models do not even permit the practical method proposed. This new method arises from a number of discrepancies between the theoretical and measured values using the current formulation of quantum mechanics. Santilli has attempted to resolve these issues by formulating what might be termed a new form of quantum mechanics, known as hadronic mechanics, which is based on a new type of mathematics called isomathematics [8]. Although abstract in nature, isomathematics has already had some definite practical success. For example, it has been used successfully to predict the growth of seashells, something which could not be done previously using conventional mathematical techniques [9]. Though only mentioned in passing, hadronic mathematics is an extensive rewrite of theory as known by most people. It is not, however, excessively complex, merely different and it is that that initially makes it hard to grasp. However, once the basic formalism is understood, much of what may be deduced follows quite straightforwardly. If this new theory is a true representation of nuclear and molecular structure, then it predicts that neutrons may be viewed as compressed hydrogen atoms. Conventionally, the probability for beta-decay of a neutron into a proton, electron and neutrino is very low for radioactive elements on a nuclear timescale; for stable isotopes, the lifetime of neutrons is effectively infinite. Hadronic mechanics predicts that such a reaction may be stimulated within the nuclei of radioactive materials.

In essence, a radioactive nucleus is in an excited energy state and is attempting to return to its ground state energy. Under normal circumstances, this is achieved by spontaneous fission or radioactive emission, the time taken to decay being dependent on how much excess energy the nucleus has. This can vary between $10^{-31}$ seconds and millions of years. An excited nucleus can return to its ground state through emission of a photon (gamma emission), an electron (beta emission), or by spontaneous fission, where alpha emission is assumed to be a form of fission. The latter two processes cause a change in the nature of the parent nucleus, altering its nuclear properties. The energy value of the excited state determines the method by which the nucleus returns to its ground state. If the decay process involves the emission of a beta particle, it may be extrapolated that a neutron will have to decay to achieve this.

From the theoretical calculations, it is hypothesised that this decay can be stimulated by bombarding the nucleus with so-called 'resonant' photons with an energy of 1.294 Mev. Under normal circumstances the probability of this interaction is extremely low. However, Santilli claims that there is a large resonance peak in the reaction cross-section (that is, the probability of the said interaction occurring) for incident photons with an energy of 1.294 Mev. It is also feasible, though not stated, that the simple existence of an excited nucleus makes it open to interaction with resonant photons, regardless of the means of decay ultimately used to return to its ground state energy. Once a neutron is converted into a proton plus reaction products, a number of possibilities could occur. Firstly, the new nucleus could be a stable isotope, in which case further interactions with the resonant photons would be unlikely and the waste would have been effectively neutralised. Secondly, the new isotope could form a new neutron deficient nucleus and one of the following could then occur:
(i) the nucleus undergoes spontaneous fission, forming two new nuclei and possibly a number of neutrons, which could interact with other fissile elements in the fuel and generate excess heat;
(ii) the neutron deficient nucleus could form a new excited energy state which can simply be categorised as another target radioactive nucleus for the resonant photons.

If this interaction is found to be true, its application for the disposal of radioactive waste is profound. Photons with the correct resonance energy can be produced easily within a piece of equipment of small volume, such that the neutraliser could be built on the same site as the parent reactor itself. Effectively, it would allow all radioactive waste to be fissioned until all the isotopes form stable nuclei. However, a point to note is that, taking a typical sample of waste, the resultant treated material would not be radioactively dangerous but chemically could be a totally unknown concoction of elements and compounds, which may well contain high levels of toxins. Another point to note is that stimulated fission would release a considerable amount of heat energy from the fuel, and so some sort of effective coolant would be required. However, since this heat energy could be used to produce even more power, there seems no reason in principle to suppose that what might be termed a secondary 'waste reactor' could not be built.

To continue quantitative scientific studies of the proposed new method for the disposal of nuclear waste essentially requires three basic experiments to be carried out. All should be of reasonable cost and are certainly realisable with present-day technology. Firstly, the experiments of Rauch and his associates [10], in which direct measurements of the alterability of the intrinsic magnetic moments of nucleons were made, should be repeated and to as high a degree of accuracy as possible. Secondly, don Borghi's experiment [11] on the apparent synthesis of the neutron from protons and electrons only should be repeated also. It is interesting to realise that, despite enormous advances in knowledge in recent times, fundamental experimental knowledge on the structure of the neutron is missing still. Finally, it is necessary to determine whether or not gamma stimulated neutron decay will occur at the resonating gamma frequency of 1.294 Mev. One way of achieving this is to have Tsagas's experiment on stimulated neutron decay [12] completed. However sceptical someone may be of these ideas, it seems sensible to perform these experiments to decide if they are valid or not. If they are valid, the rewards would be tremendous.

Even assuming that the theory is found to be sound and the predicted resonance peak exists, there would still be further practical considerations when applied to the disposal of radioactive waste. Nevertheless, it is easy to see that, if proven, such a method would save a truly considerable amount of public funds, given the relatively low cost of the apparatus as compared with the removal of the need to transport the spent fuel to reprocessing facilities and also with the building of long-term storage facilities. The possibility of producing toxic by-products is, however, a real concern and a means for the disposal of such by-products, if they did materialise, would have to be sought as a matter of urgency.

**Conclusion.**

Hence, the world faces an almost exponentially increasing demand for energy due to the underdeveloped sections of the world becoming more industrialised and demanding an improved standard of living and this position is exacerbated by the rapid increase in the worlds' population. This ignores the possibility of a further increase in demand due to the introduction of new technology. This demand cannot be met by the use of fossil fuels and, in any case, if it could, the increased use of such fuels would surely have a less than beneficial effect on the environment. The regenerative and so-called 'renewable' forms of energy production are seen to be able to make a contribution, particularly locally, but they do not seem capable of having a truly major effect. Although not mentioned previously, it may be noted that the constructing of a first nuclear fusion reactor seems as far away as ever; indeed, many feel such a reactor impossible to build. It seems, therefore, that, with the existing state of human knowledge, the only viable energy source sufficient for supplying the future energy needs of the world is nuclear power. It has to be recognised that there are attendant problems. People are, and probably will be for a long time, very uneasy about nuclear power. They've seen its awful potential destructive power and so, quite naturally, worry about the possibility of accidents, even catastrophic accidents, at the plants themselves. People are also very well aware of the major problem posed by nuclear waste. Although the traditional methods of dealing with this waste are acceptable, they are politically controversial and/or extremely expensive in monetary terms, both factors being highly important in the case of the location of underground storage facilities. Various others methods have been advocated over the years but not one has remained in favour for long. Here attention has been drawn to the relatively new ideas proposed by Santilli. They are revolutionary in concept, they do draw on a new form of mathematics and quantum mechanics but tests have been carried out already to see if the theory works. More tests are being carried out but the initial results are positive. If the ideas are eventually proven, they will provide the possibility for a means of radioactive waste disposal which satifies the requirements for convenience, finality of disposal, political acceptance and cost. As with all new ideas there is scepticism within the existing scientific community but, if Santilli's theories are finally supported by experimental evidence, few grounds for objection could remain for what could be a revolutionary technology. It is to be hoped that experimentation to validate, or otherwise, Santilli's theories will continue.

**References.**

[1] Population Reference Bureau - World population datasheet 2002.

[2] Cole, G.H.A., in *Entropy and Entropy Generation*, J.S.Shiner(ed), pp159-173, Kluwer Acad. Pub., Netherlands, 1996.

[3] Bell, A., *Energy 1 - Fossil Fuels*, p125, Open Univ., 1995

[4] www.eia.doe.gov/emeo/ipsr/t21.xls

[5] Scott, M. & Johnson, D., *Nuclear Power* Open Univ., 1993

[6] *National Gepgraphic*, July 2002

[7] *Access to Energy*, September 2002

[8] Santilli, R.M., *Foundations of Hadronic Chemistry* and references cited there. Kluwer Academic Publishers, Dordrecht, 2001

[9] Santilli, R.M., *Isotopic, genotopic and hyperstructural methods in theoretical biology* Naukova Dumka Publisher, Ukraine, 1996

[10] Rauch, H. Et al., Phys. Lett. A **54**(1975)425

[11] Borghi, C., J. Nuclear Phys. **56**(1993)147 (in Russian)

[12] Tsagas, N.F., Hadronic J. **19**(1996)87